\documentclass[pdflatex,iicol,sn-mathphys-num]{sn-jnl}

\usepackage{graphicx}%
\usepackage{multirow}%
\usepackage{amsmath,amssymb,amsfonts}%
\usepackage{amsthm}%
\usepackage{mathrsfs}%
\usepackage[title]{appendix}%
\usepackage{xcolor}%
\usepackage{textcomp}%
\usepackage{manyfoot}%
\usepackage{booktabs}%
\usepackage{algorithm}%
\usepackage{algorithmicx}%
\usepackage{algpseudocode}%
\usepackage{listings}%
\usepackage{siunitx}%
\usepackage[capitalise]{cleveref}%
\usepackage{orcidlink}%

\begin{document}

\title{GPT-like transformer model for silicon tracking detector simulation}


\author*[1]{\fnm{Tadej} \sur{Novak} \orcidlink{0000-0002-3053-0913}}\email{tadej.novak@cern.ch}  
\author[1,2]{\fnm{Borut Paul} \sur{Ker\v{s}evan} \orcidlink{0000-0002-4529-452X}}  
\affil[1]{
  \orgdiv{Experimental particle physics department},
  \orgname{Jo\v{z}ef Stefan Institute},
  \orgaddress{\street{Jamova cesta 39},
    \postcode{1000} \city{Ljubljana},
    \country{Slovenia}}%
}
\affil[2]{
  \orgdiv{Faculty of Mathematics and Physics},
  \orgname{University of Ljubljana},
  \orgaddress{\street{Jadranska ulica 19},
    \postcode{1000} \city{Ljubljana},
    \country{Slovenia}}%
}

\abstract{Simulating physics processes and detector responses is essential
  in high energy physics and represents significant computing costs.
  Generative machine learning has been demonstrated to be potentially powerful
  in accelerating simulations, outperforming traditional fast simulation
  methods. The efforts have focused primarily on calorimeters.
  This work presents the very first studies on using neural networks for
  silicon tracking detectors simulation.
  The GPT-like transformer architecture is determined to be optimal for this
  task and applied in a fully
  generative way, ensuring full correlations between individual hits.
  Taking parallels from text generation, hits are represented as a flat sequence
  of feature values. The resulting tracking performance, evaluated
  on the Open Data Detector, is comparable with the full simulation.}


\maketitle

\section{Introduction}\label{sec:intro}

A large part of the physics programme of the Large Hadron Collider (LHC) relies
on accurate simulation of collision events that complement the collected data.
Traditionally, simulations rely on Monte Carlo (MC) methods, which are highly
accurate but also account for a significant portion of the experiments'
computing requirements~\cite{CERN-LHCC-2022-007,HEPSoftwareFoundation:2017ggl}.
Monte Carlo simulation can be divided into four main steps~\cite{ATLAS:2010arf}:
generation of physics events and the immediate particle decays,
simulation of the detector and particle interactions,
digitisation of the energy and charge deposited on the sensitive regions
of the detector into formats comparable with the actual detector readout,
and the subsequent reconstruction using the same algorithms
as for data collected by the experiments.

The detector response simulation part is computationally the most
demanding~\cite{HEPSoftwareFoundation:2018fmg},
especially the simulation of particle showers in calorimeter systems.
It is commonly performed using the \textsc{Geant4} simulation toolkit~\cite{GEANT4:2002zbu}.
In recent years, generative machine learning (ML) models have emerged
as a proposed solution to significantly accelerate simulation speed while
keeping physics performance as close to \textsc{Geant4} as possible,
focusing on calorimeters~\cite{Krause:2024avx}.
As an additional benefit, they can also use the accelerated processors
(e.g. GPUs) for increased speed compared to standard computer processors (CPUs)
out of the box, i.e.\ without dedicated code rewrites.
Various different techniques have been applied, including generative adversarial
networks (GANs)~\cite{Paganini:2017hrr,Paganini:2017dwg,Erdmann:2018kuh,Erdmann:2018jxd,Musella:2018rdi,Belayneh:2019vyx,Butter:2020qhk,ATLAS:2021pzo,ATLAS:2022jhk,Hashemi:2023ruu,FaucciGiannelli:2023fow,Simsek:2024zhj},
variational autoencoders (VAEs)~\cite{Buhmann:2020pmy,Buhmann:2021caf,Cresswell:2022tof,Bieringer:2022cbs,Diefenbacher:2023prl,Hoque:2023zjt,Liu:2024kvv,Smith:2024lxz},
classical normalising flows~\cite{Krause:2021ilc,Krause:2021wez,Krause:2022jna,Diefenbacher:2023vsw,Xu:2023xdc,Buckley:2023daw,Pang:2023wfx,Ernst:2023qvn,Schnake:2024mip,Du:2024gbp,Buss:2024orz},
autoregressive models~\cite{Birk:2025wai},
diffusion and continuous flow models~\cite{Mikuni:2022xry,Buhmann:2023bwk,Acosta:2023zik,Mikuni:2023tqg,Amram:2023onf,Buhmann:2023kdg,Jiang:2025pil,Kobylianskii:2024ijw,Jiang:2024bwr,Favaro:2024rle,Brehmer:2024yqw,Buss:2025cyw,Raikwar:2025fky},
and combinations of different model types~\cite{Favaro:2025ift,Buss:2025kiu}.
For a recent taxonomy see e.g. Ref.~\cite{Hashemi:2023rgo}.
The ATLAS experiment developed a realistic setup, where GANs are
already used in production~\cite{ATLAS:2021pzo}, with normalising flows
and diffusion models being evaluated for the next generation~\cite{ATL-SOFT-PUB-2025-003}.

However, by the end of the High-Luminosity LHC programme (HL-LHC)~\cite{ZurbanoFernandez:2020cco},
the ATLAS and CMS experiments are expected to have collected up to ten times
the amount of data recorded during the first three runs.
As a result, many of the cutting-edge physics analyses will start to have their
sensitivity limited by the available statistics of the Monte Carlo simulation.
A simulated sample larger than the collected data will be required to increase
the precision of Standard Model background modelling.
To match the expected resource constraints, the goal is to produce
significant fractions of the MC samples using fast simulation methods.
The available machine learning tools will have to be extended to cover
the whole detector, including silicon tracking detectors.

\begin{figure}[t]
  \includegraphics[width=0.8\columnwidth]{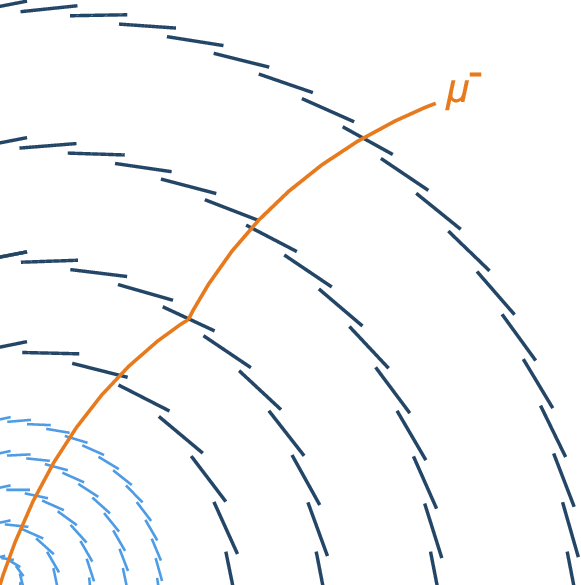}
  \caption{A schematic illustration of a muon track in a two-component silicon
    tracking detector system in a transverse plane.
    The light and dark blue lines represent sensitive detectors of two types.
    The orange curved line represents a muon track that scatters on
    a detector element}%
  \label{fig:sketch}
\end{figure}

This study explores the first use of generative machine learning for
the simulation of silicon tracking detectors.
For traditional calorimeters, the simulated detector response can be interpreted as an image
with correlated entries. On the other hand, in tracking detectors each track
is independent, described by a sequence of interactions with the detector,
also called hits. Similar holds also for calorimeter simulation of highly-granular
detectors. Such response can only be described with a very empty
image as sensitive detectors are thin and relatively far apart, as sketched
in \cref{fig:sketch}. This motivates a sequence-based machine learning model.

The original idea of this paper is to encode the simulated track information
in a sequence inspired by large language models. Like in a generative
pre-trained transformer (GPT) the task of the neural network is to predict
consecutive hits in the series. This ensures correlations between the input
hits and the predicted hit that follows. Hit properties need to be represented
in a discreet token space which is challenging for continuous numerical
features.
Consequently the state of the art large language models are well suited
for this problem without the need for substantial adaptations.
The challenge thus lies in finding an optimal way to prepare and encode
the data.

The rest of the paper is organised as follows.\@
\cref{sec:odd} briefly introduces the Open Data Detector used for this study,
\cref{sec:datasets} contains the overview of the datasets used, how they are
simulated and prepared for machine learning usage.\@
\cref{sec:model} presents a detailed description of the transformer model used,
both the model itself and the training setup.\@
\cref{sec:results} describes the validation procedure and presents the results
for single muon, electron and pion samples.
Finally, conclusions are presented in \cref{sec:conclusions}.

\section{The Open Data Detector}\label{sec:odd}

The Open Data Detector (ODD)~\cite{Gessinger-Befurt:2023snx}
is a generic, HL-LHC style tracking detector, that is loosely modelled after
the ATLAS Inner Tracker (ITk)~\cite{ATLAS-TDR-30,ATLAS-TDR-25}.
It comprises a pixel system, a short- and a long-strip system
and provides a simplified but realistic detector in terms of geometric layout,
number and type of sensitive elements, and passive material.

The ODD pixel system consists of four cylindrical barrel layers of sensors,
accompanied by seven endcap disk layers on each side. To achieve maximum
coverage sensors are staggered in both the azimuthal and longitudinal
direction. In total 3332 pixel sensors are part of the ODD.\@

Silicon strip sensors are split in two categories based on how they
are segmented. Long strips refer to strips stretching across the full sensor,
while short strips have multiple strip segments in the strip direction.
The former are typically mounted in pairs, rotated with respect to one another,
to provide a two-dimensional measurement of a particle intersection.
The short strip system of the ODD consists of four cylindrical barrel layers
and six endcap disks on either side, while the long strip system consists
of two barrel layers and six endcap disks. Together they comprise 9714 detector
modules.

The ODD tracker system also encompasses a solenoid of \qty{1.2}{\metre} radius,
which produces a magnetic field of \qty{2.6}{\tesla} in the center
of the detector, which enables momentum measurement through bending
of charged particle trajectories. Calorimeters are also part of the Open Data
Detector but are not used for this study. The geometry is integrated
in the ACTS software suite~\cite{Ai:2021ghi} allowing detailed detector
simulation using \textsc{Geant4}.

\section{Datasets}%
\label{sec:datasets}

Four single particle samples are used for machine learning model training,
validation and performance evaluation, and are generated using the setup
described in the previous section. To keep the phase-space small
selection on particle transverse momentum $p_\mathrm{T}$, pseudorapidity $\eta$
and the polar angle $\phi$ is applied. Three benchmark datasets are generated
for $\mu^{-}$, $e^{-}$ and $\pi^{+}$ with
$\qty{80}{\GeV} < p_\mathrm{T} < \qty{85}{\GeV}$,
$0.05 < \eta < 0.1$ and $0 < \phi < 0.1$, one million events each.
In addition, a larger, 100 million event sample with both muons and anti-muons
is produced with loosened selection
$\qty{70}{\GeV} < p_\mathrm{T} < \qty{90}{\GeV}$,
$0.05 < \eta < 0.25$ and inclusive in $\phi$.
In the ATLAS fast calorimeter simulation production setup models are also
sliced in $p_\mathrm{T}$ and $\eta$, but inclusive in $\phi$, so this
sample represents a realistic dataset that could eventually be used
in a production setup.
A summary of the datasets used is shown in \cref{tab:datasets}.
Particles are generated at the detector center where the beamspot position
is smeared in the $z$ coordinate uniformly between
-50 and \qty{50}{\milli\metre}.

\begin{table}[t]
  \caption{Summary of the datasets used in this study. Single particle samples
    are simulated in a subset of $p_\mathrm{T}$ range (in GeV),
    $\eta$ and $\phi$.
    A single charge is simulated with the exception
    of the larger single $\mu^{\pm}$ sample}\label{tab:datasets}%
  \begin{tabular*}{\columnwidth}{@{\extracolsep\fill}llllr@{\extracolsep\fill}}  
    \toprule
    Dataset            & $p_\mathrm{T}$ & $\eta$     & $\phi$ & \multicolumn{1}{l}{Events} \\
    \midrule
    single $\mu^{-}$   & 80--85         & 0.05--0.1  & 0--0.1 & \num{1000000}              \\
    single $\mu^{\pm}$ & 70--90         & 0.05--0.25 & incl.  & \num{100000000}            \\  
    single $e^{-}$     & 80--85         & 0.05--0.1  & 0--0.1 & \num{1000000}              \\
    single $\pi^{+}$   & 80--85         & 0.05--0.1  & 0--0.1 & \num{1000000}              \\
    \bottomrule
  \end{tabular*}
\end{table}

The particle interactions with the detector are simulated using \textsc{Geant4}.
Each interaction, with either sensitive or passive material, is called a hit.
One of the following processes may occur at each hit:
multiple-scattering, energy loss due to ionisation and radiation
(bremsstrahlung), photon conversion (electron-positron pair creation)
or hadronic interaction.

Standard detector simulations only consider hits in sensitive detector
material volumes.
They are later digitised into digital signal analogous to the actual detectors.
While interactions with the passive material are important for realistic
particle trajectory they do not need to be stored.
For simplicity and to keep the simulation output representation the same
only hits occurring in the sensitive detectors are considered also for this
study. This allows to utilise the layered structure of the detector and describe
the simulated detector response to an individual particle as a sequence
of discrete hits.
While secondary particles are allowed to be produced in the simulation,
they are to be treated independently and are not considered for this study.
The secondary hits they produce are discarded.

Each hit can be described with 7 primary features.
Firstly, particle type and charge are encoded in a discrete particle ID.\@
Each sensitive detector module is also assigned its own geometry ID.\@
No additional readout segmenting is performed,
meaning each detector is treated as a continuous block of silicon.
The hit position can be described either in terms of global or local
coordinates. To ensure hits occur in the actual sensitive detector regions
and not in vacuum, each hit position is projected onto a corresponding
tracking surface. As the thickness of the detector is much smaller than
the surface area, it is ignored and the final transformation encodes
the position in a 2D space yielding two local continuous coordinates.
Such a projection also ensures a finite range of allowed coordinate values
that only depends on detector type and not on the global position
of the detector module.
Finally the particle momentum after the hit provides three additional
continuous features.

\begin{figure}[htbp]
  \centering
  \includegraphics[width=\columnwidth]{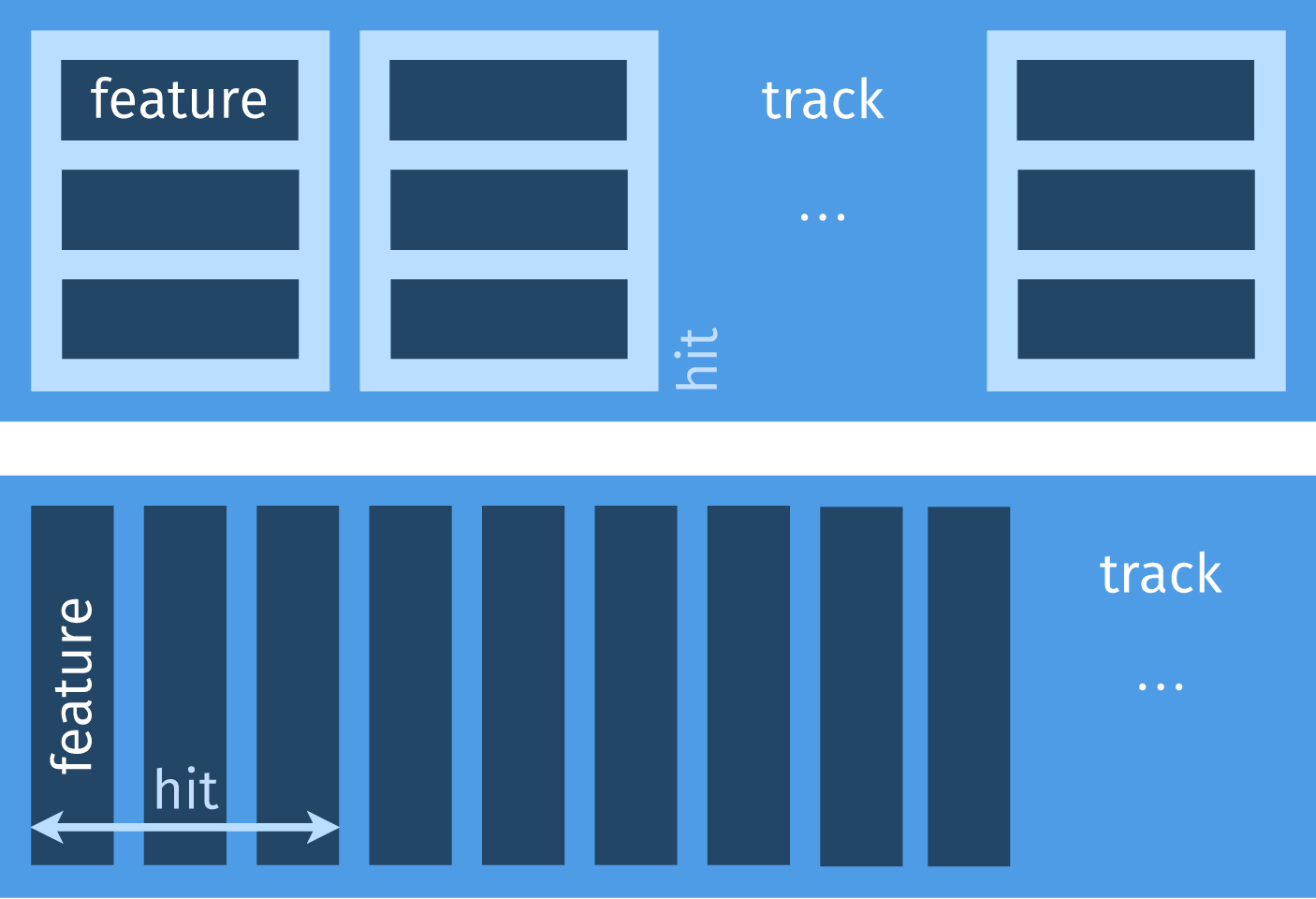}
  \caption{Graphical illustration of the track hits data representation
    in a 3D way, as in the output of the simulation (top)
    and a 2D flattened way, as used in the transformer model (bottom).
    For simplicity three features per hit are used in the illustration}%
  \label{fig:representation}
\end{figure}

To describe a collision event, each particle track is assigned its own sequence
of hits starting from the interaction point. An additional start hit is defined
using a virtual module with the initial momentum and the beamspot position.
The sequence ends once the particle leaves the tracking detector,
also described with an additional virtual hit. An illustration of such
3D representation is presented in \cref{fig:representation}.
Alternatively hit information inside a track could be flattened in a sequence
of features with a deterministic order yielding a more transformer-friendly
2D representation of a particle track.

\section{The Transformer Model}%
\label{sec:model}

A transformer~\cite{Vaswani:2017lxt} is a deep neural network architecture
used for sequence modelling and transformation, most commonly for natural
language processing tasks.

First, text or other discrete information is split into tokens and assigned
a sequential numerical representation. As part of the training, every token
gets assigned a vector in the embedding space.
The machine learning model itself is built using two main components,
the encoder and the decoder. Both contain multiple layers, each containing
the self-attention part and the feed-forward neural network.
The former determines the relative importance of each sequence element relative
to the others and the latter acts as the traditional fully-connected neural
network layer.
In transformers, multiple attention layers run in parallel, a mechanism known
as multi-head attention.
Since the model contains no recurrence or convolution, a sinusoidal positional
encoding~\cite{Vaswani:2017lxt} is applied to add sequence order information
at embedding level in cases where sequence order is important.

When simulating the silicon detector response, the goal of the model
is to predict the next element of the particle track sequence, where both input
and output work on the same data representation.
A decoder-only architecture is used in this case.
As an optimisation, all sub-sequences are modelled at the same
time using masking, where all sequence elements after the current position
are not taken into account in the self-attention mechanism.

Appropriate tokenisation procedure is crucial for efficient transformer model
training.
Every feature is tokenised separately but sharing a common dictionary with
token space consisting of sequential integers starting with zero, which
is reserved for the padding and end token.
First, every discrete feature is tokenised one by one. Each of them is assigned
a unique set of tokens.
At the end all continuous features are tokenised together and share the same
tokens at the end of the token space. As a finite representation is required,
continuous features are additionally rounded to two decimal places.
All unique numerical values from the whole dataset get assigned a token.
By describing the hit position using the local surface coordinate no additional
standardisation is needed. Given that momenta values take similar ranges
no additional standardisation is applied there to keep the same precision
for all continuous features. 
This procedure is inspired by Ref.~\cite{nagy2023generative}.
To reduce the token space size numerical features could be further split
by number of digits, but this makes the sequence longer and increases
the inference complexity.

The 2D representation of particle tracks is used yielding a flat token array
where each feature represents a single individual element of the sequence.
At the inference stage the usage of deterministic ordering allows sampling only
the allowed set of tokens per feature, which is stored as part of
the dictionary.

The output of the transformer is passed through a final fully-connected
layer, where the output vector dimension equals to the token dictionary
size. When normalised to unity the output of the model can be interpreted
as probability for a token to occur at the next position in the sequence.
A cross-entropy loss is computed between the model output and the true token
at the specific sequence position.

The \texttt{nanoGPT}~\cite{nanoGPT} implementation of a decoder-only transformer
is used. Eight layers with eight heads each are used in all of the models
tested. Feed-forward dimension in each of the layers is four times larger
than the input dimension. Two different input dimensions are tested,
256 and 512, as models trained on larger token space perform better with
larger input dimensions. 
Dropout of 0.2 is applied at the end of each of the layers.
The parameters of the model configurations used are specified
in detail in \cref{tab:model}.
As the model size depends on the token dictionary size,
the final tested model sizes range between 9.1 and 35 million parameters.

\begin{figure}[htbp]
  \centering
  \includegraphics[width=\columnwidth]{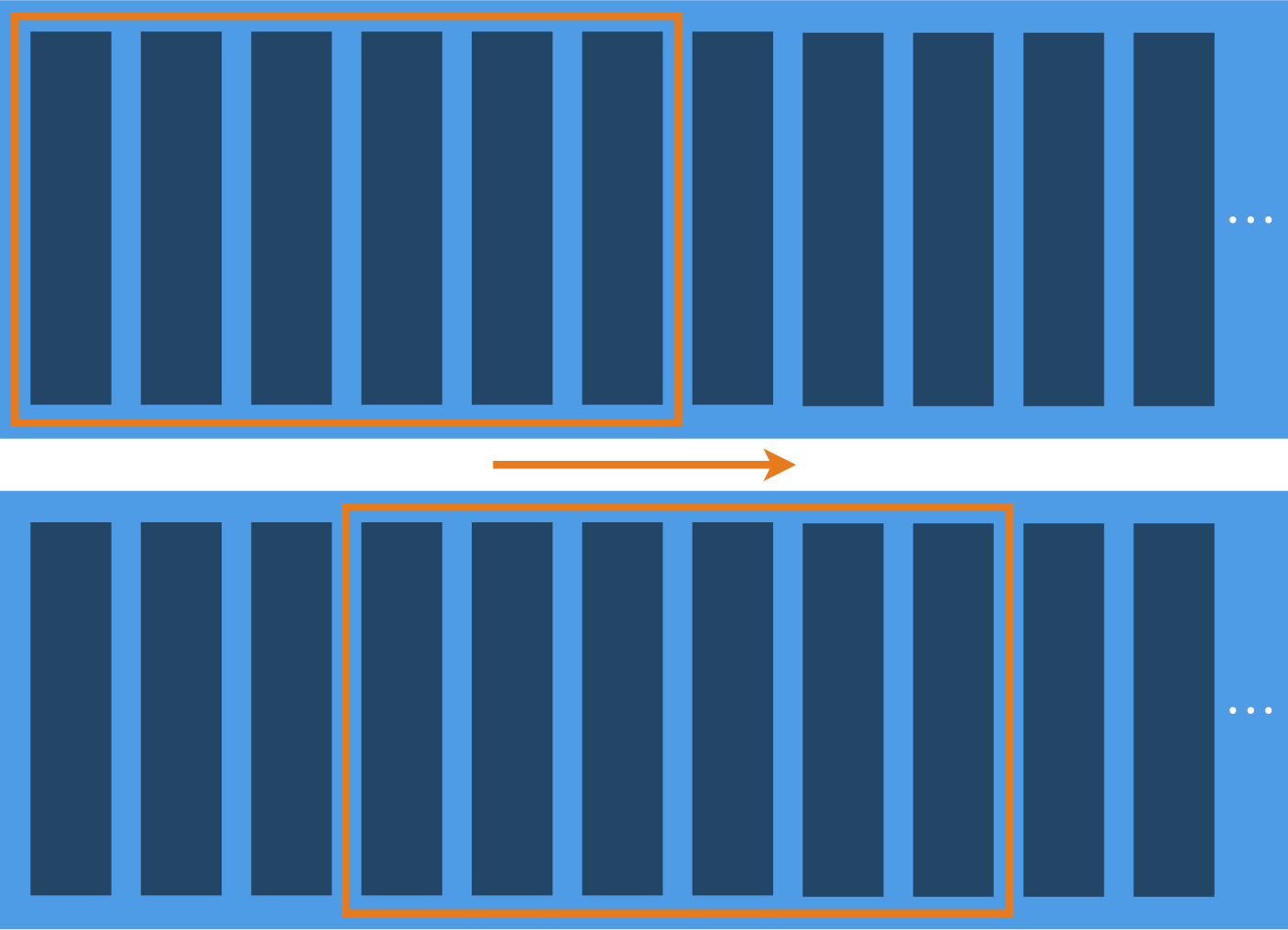}
  \caption{Illustration of the sliding window attention training.
    Individual full sequences are split in sliding windows of multiple hits.
    Models are then trained on individual windows.
    For simplicity three features per hit and two hits per window are used
    in the illustration}\label{fig:windowed}
\end{figure}

The attention mechanism is the slowest part of the model, having a quadratic
complexity dependence on the sequence size. To reduce the size of the sequence
the sliding windowed attention training is performed.
Particle track sequences are split in rolling windows of four hits, e.g.\ for
a sequence of 20 hits this yields 17 smaller sequences.
This is acceptable from the physics perspective because the correlation between
hits drops with their distance, the most important physics feature being the
(local) track curvature.
While this procedure augments the training dataset, making it an order
of magnitude larger, the graphics accelerators cope much better with larger
batch sizes.
An additional auxiliary feature is added to each hit representing the hit index
in the sequence. Maximum sequence size is thus 36 in all the tested scenarios.
The sliding window attention training is illustrated in \cref{fig:windowed}.

\begin{table*}[htbp]
  \centering
  \caption{Transformer model parameters used. The core of the model is constant
    and only the dimensions change. Final model size depends on the token
    dictionary size}%
  \label{tab:model}
  \begin{tabular*}{\textwidth}{@{\extracolsep\fill}lccccc}  
    \toprule
    & single $\mu^{-}$ & single $\mu^{\pm}$ & single $\mu^{\pm}$ (large) & single $e^{-}$ & single $\pi^{+}$ \\
    \midrule
    input dimension        & 256              & 256                & 512                        & 256            & 256              \\
    feed-forward dimension & 1024             & 1024               & 2048                       & 1024           & 1024             \\
    layers                 & 8                & 8                  & 8                          & 8              & 8                \\
    heads                  & 8                & 8                  & 8                          & 8              & 8                \\
    dropout                & 0.2              & 0.2                & 0.2                        & 0.2            & 0.2              \\
    \midrule
    max.\ sequence length  & 36               & 36                 & 36                         & 36             & 36               \\
    token dictionary size  & 11301            & 19125              & 19125                      & 15942          & 11099            \\
    categorical tokens     & 70               & 2245               & 2245                       & 2858           & 71               \\
    total model parameters & \qty{9.2}{M}     & \qty{11.2}{M}      & \qty{35.0}{M}              & \qty{10.4}{M}  & \qty{9.1}{M}     \\
    \bottomrule
  \end{tabular*}
\end{table*}

Training is performed using the \texttt{AdamW} optimiser~\cite{Loshchilov:2017bsp},
an enhanced version of \texttt{Adam}, where weight decay is applied during
the parameter update, leading to more consistent regularisation and better
generalisation.
Additionally gradient clipping is enabled to prevent exploding gradients
affecting the result too significantly. This is achieved by limiting the
maximum value of the gradient norm to 5.0.
The learning rate is varied for a factor of 10 between the minimum and maximum
value. It starts at the minimum value and linearly rises to the maximum value
for three epochs. Then it decays back to the minimum over a cosine half-period
for 500 epochs. The remainder of the training uses a constant learning rate.

Input datasets are split into training, validation and test samples in
the 8 : 1 : 1 ratio. The training part is used for the training while  
the validation sample is used to track the performance of the model after
each training epoch.
Graphics accelerators are used for training at two different precision levels,
the single 32-bit floating-point precision (\texttt{fp32}), and the half
16-bit precision (\texttt{bf16}), where 8 exponent bits are retained.
Training batch size depends on the available memory and is picked at
\num{3500} for \qty{40}{GB} of memory and scaled linearly for larger available
memory amounts.
The minimum learning rate is chosen to be \num{1e-4} for this batch size
and is also linearly scaled with increasing batch sizes to keep the model
training rate the same on different GPUs.
The trained model yielding the lowest validation loss is taken,
reached at the order of \num{5000} training epochs.

Inference runs on the starting virtual hit representing the initial conditions
of the particle, where no detector-level information is assumed.
The iterative process predicts the next feature at each step for the whole
batch, sampling only from the allowed token set. No additional filtering
or scaling based on probabilities is applied. Due to the sliding window
attention setup at most three hits are used as input to the inference at
a given step. The iteration stops when all tracks in the batch leave
the detector and reach the end token.

The training and inference code is collected in the \texttt{SiliconAI}
software package~\cite{SiliconAI}.

\section{Validation and Results}\label{sec:results}

\begin{figure}[b]
  \centering
  \includegraphics[width=0.9\columnwidth]{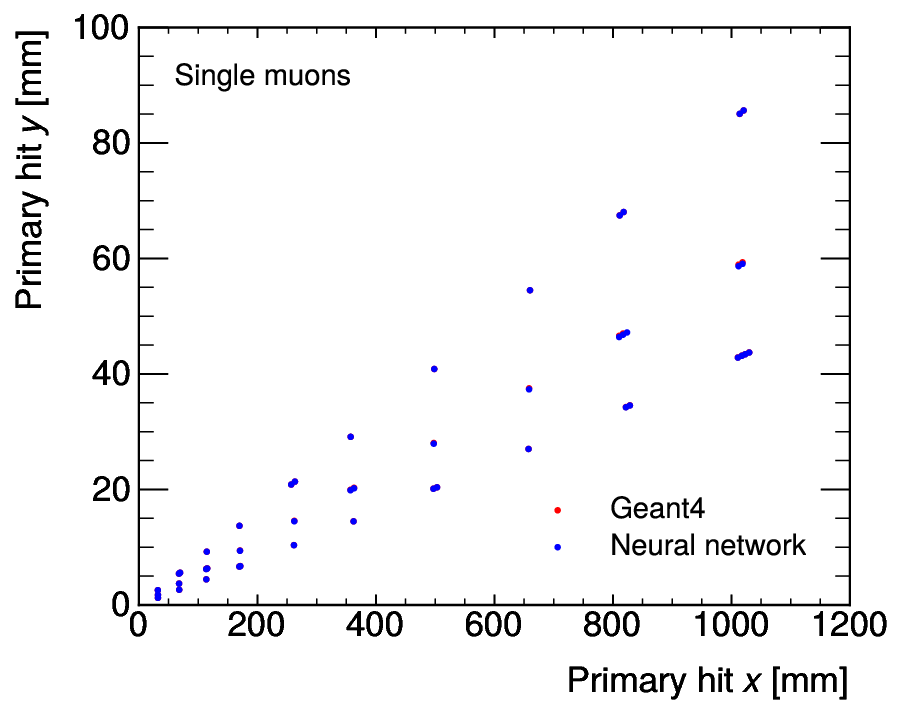}
  \caption{Hits produced by three representative $\mu^{-}$ trajectories,
    particles simulated with \textsc{Geant4} (red) and the neural
    network (blue).}\label{fig:results:scatter:mu}
\end{figure}

As introduced in \cref{sec:model}, five different models are trained.
The inference is performed on the test sample to ensure that only events
never seen by the neural network are used for validation.
Two levels of validation are performed. Hit-level validation compares
cumulative distributions of hit properties with the rounded reference
\textsc{Geant4} simulation, on which the model was also trained on.
Three representative muon trajectories as predicted by the two simulation
types are shown in \cref{fig:results:scatter:mu}.
The goal of this paper is to evaluate how well neural networks can be trained
to generate silicon tracker hits, so results will always be compared with
rounded \textsc{Geant4} data, even if that does not provide the same
physics performance as the unmodified \textsc{Geant4} hits.

Higher-level track-based validation is also performed. Simulated hits
are reconstructed using the ACTS tracking software using the default Open Data
Detector reference configuration~\cite{SiliconAI_Validator}.
The performance of each simulation model is evaluated at the track seeding
stage and at the very end of the track reconstruction workflow.

Technical efficiencies will be used in this paper. Compared to the physics
efficiency, which is the efficiency for a given generated particle to get
a track reconstructed, the technical efficiency does not depend on the detector
material or on the detector coverage, allowing the isolation of the algorithmic
efficiency. The technical seeding efficiency represents the efficiency to find
seeds for reconstructable particles, and is defined as the fraction of seed
matches among particles providing at least three measurements in the detector.
The technical tracking efficiency represents the fraction of track matches among
the charged particles providing enough measurements in the detector to satisfy
the reconstruction cuts.

Seeding efficiency is a good initial estimate of the quality of simulated
particle trajectories, including their compatibility with
a helix trajectory associated with a particle with the estimated momentum.
Low seeding efficiency would mean that many tracks do not satisfy the maximum
allowed multiple-scattering effect. While this is a parameter of the tracking
that can be tuned, the goal is that the same tracking setup would be used
for \textsc{Geant4} and the transformer model.

Finally the final reconstructed tracks should match well the ones resulting
from the standard simulation. Both track properties directly and their
associated resolutions are compared. A pull of a track property can be defined
as the difference between the reconstructed value and the true value, divided
by the resolution of the property.

\subsection{Single Muons}

\begin{figure*}[htbp]
  \centering
  \includegraphics[width=0.9\columnwidth]{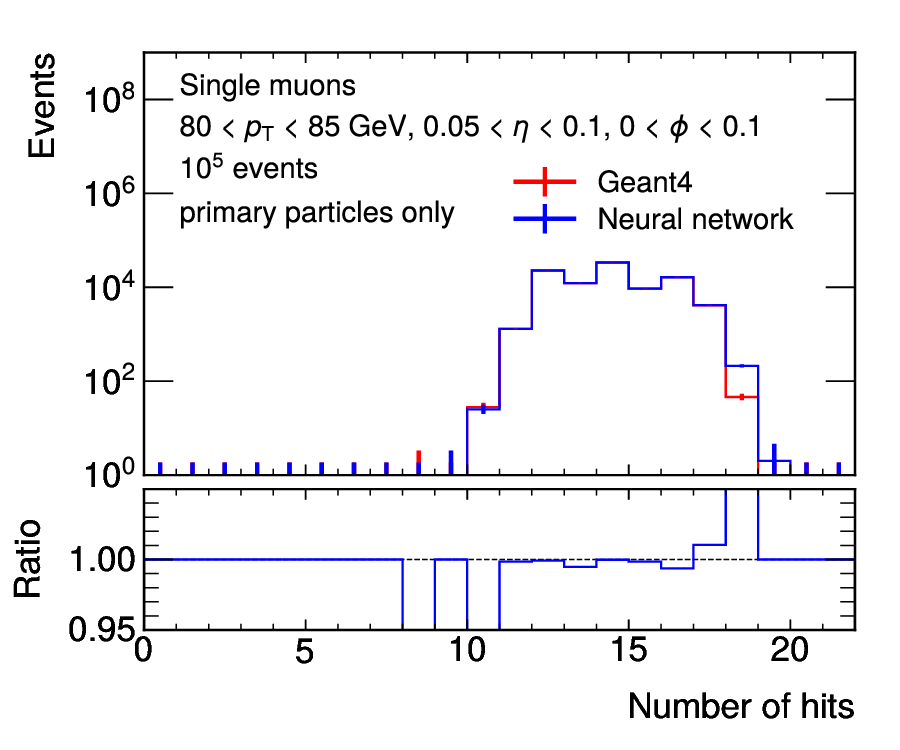}
  \quad
  \includegraphics[width=0.9\columnwidth]{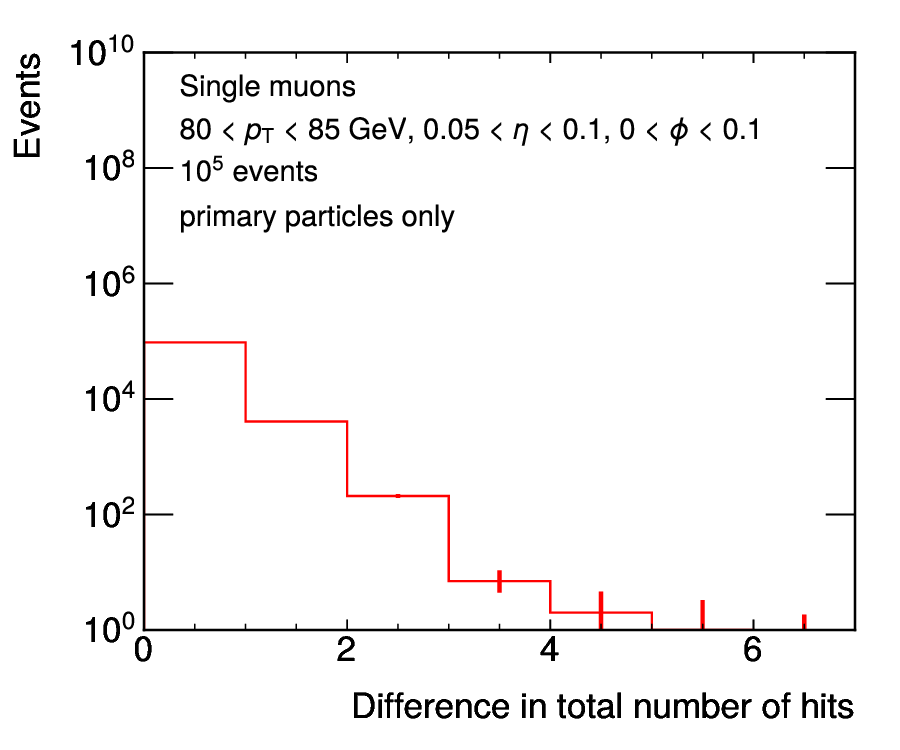}
  \caption{Comparisons of numbers of simulated hits of single $\mu^{-}$
    particles simulated with \textsc{Geant4} (red) and the neural
    network (blue).
    The total number of hits (left) and the difference
    in total number of hits (right) are shown}\label{fig:results:nhits:mu}
\end{figure*}

\begin{figure*}[htbp]
  \centering
  \includegraphics[width=0.9\columnwidth]{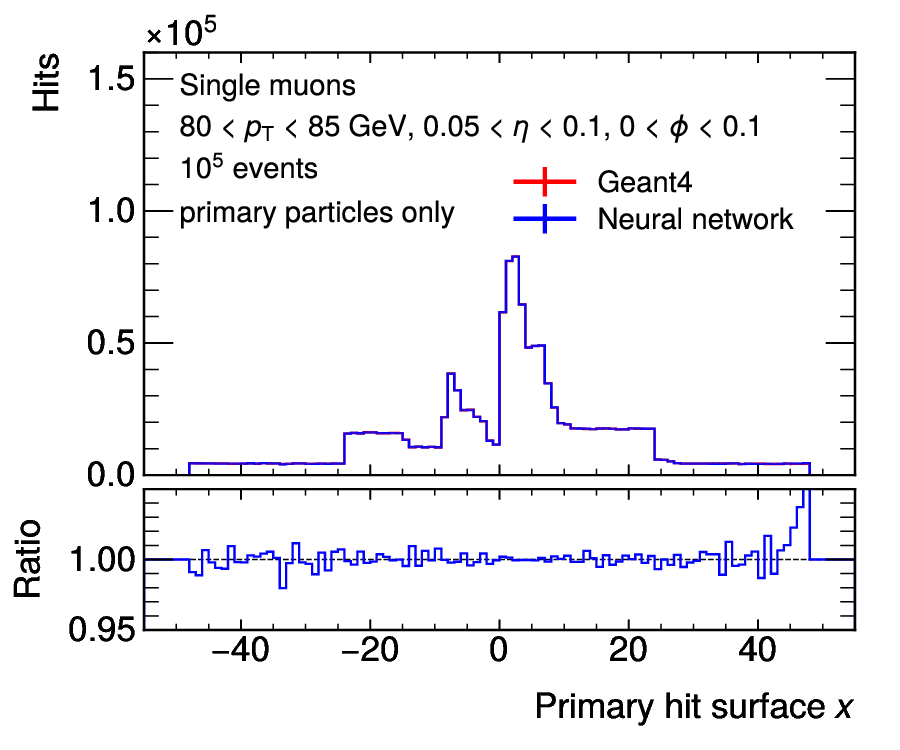}
  \quad
  \includegraphics[width=0.9\columnwidth]{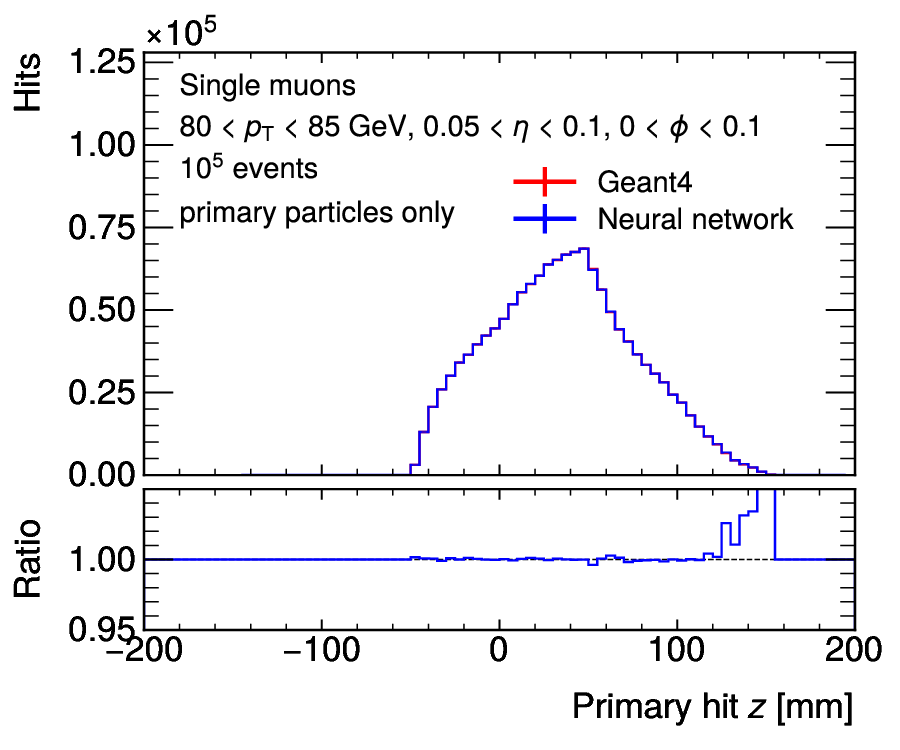}
  \caption{Comparisons of simulated hit properties of single $\mu^{-}$
    particles simulated with \textsc{Geant4} (red) and the neural
    network (blue).
    Surface coordinate $x$ of the hit (left) and
    global hit coordinate $z$ (right) are shown}\label{fig:results:hits:mu}
\end{figure*}

\begin{figure*}[htbp]
  \centering
  \includegraphics[width=0.9\columnwidth]{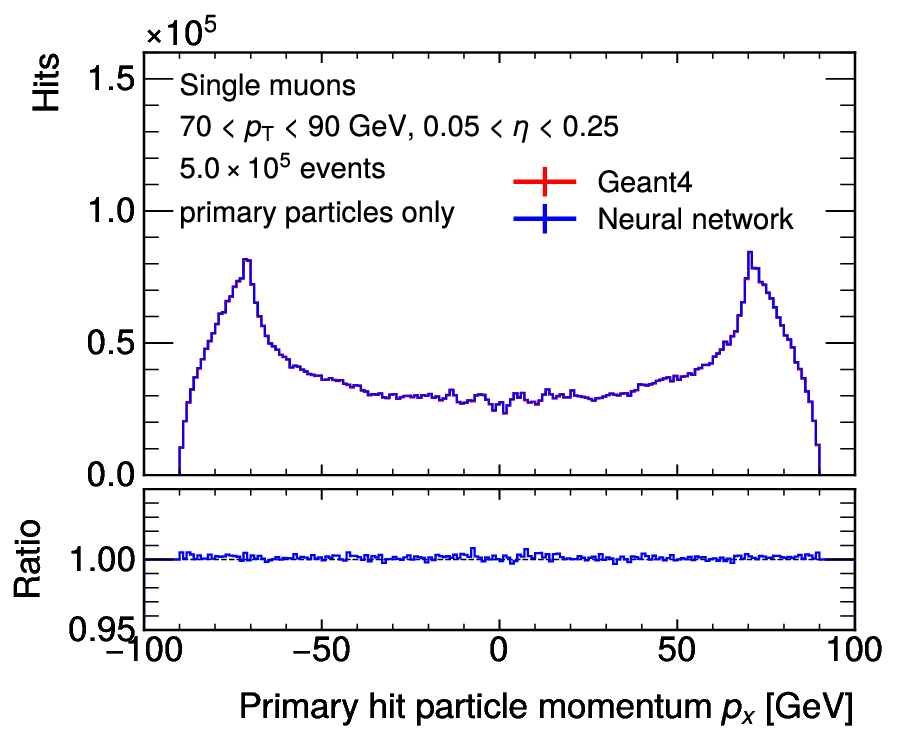}
  \quad
  \includegraphics[width=0.9\columnwidth]{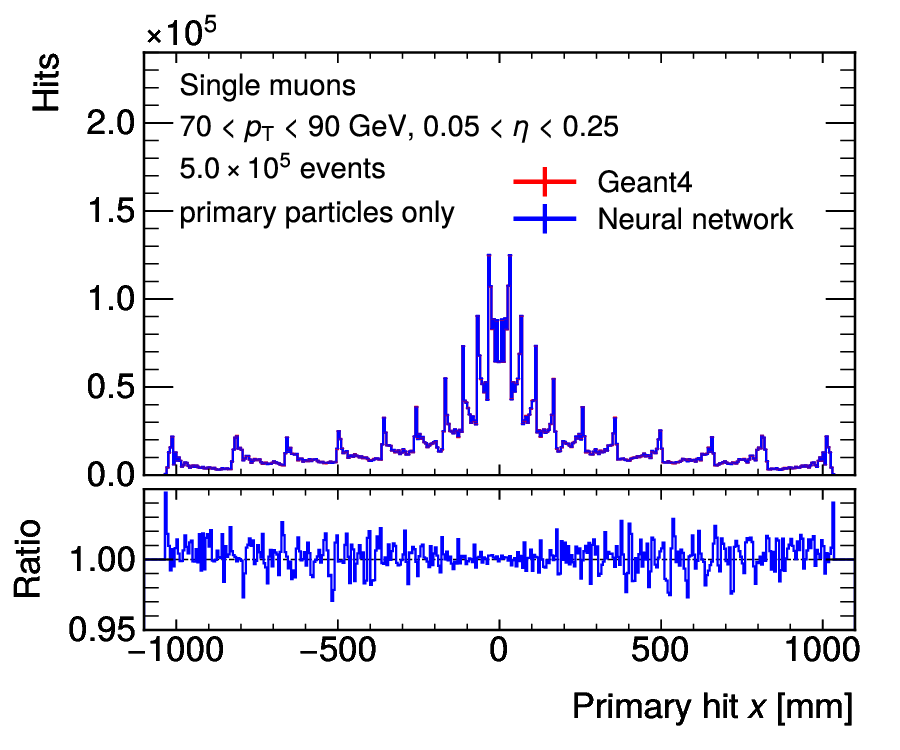}
  \caption{Comparisons of simulated hit properties of single $\mu^{\pm}$
    particles simulated with \textsc{Geant4} (red) and the neural
    network (blue).
    Particle hit momentum $p_{x}$ at the hit (left) and global hit coordinate
    $x$ (right) are shown}\label{fig:results:hits:mu_large}
\end{figure*}

Single muons, due to their physics nature, interact with the detector
the least among the tested datasets.
The multiple-scattering occurs in the vast majority of the cases
and almost no momentum loss is expected.
Muons are a good starting point to evaluate the performance of the models.

Three different models are trained for single muon particles, a \qty{9.2}{M}
parameter model with only muons and smaller phase-space, and two
models with different input dimension size using both muons and anti-muons
in a larger detector coverage.
\Cref{fig:results:nhits:mu} shows the number of simulated hits for the smaller
($\mu^{-}$ only) dataset. Number of hits is well modelled with slightly larger
deviations for high number of hits where two orders of magnitude less events
are expected.
The majority of events yield the same number of hits for both simulation
types, followed by about an order of magnitude drop with increasing
deviation. The largest hit number difference for any of the models trained
for muons is six.

The hit distributions in coordinates and momentum agreement is also very good,
as shown on \Cref{fig:results:hits:mu,fig:results:hits:mu_large}.
Hit properties that are directly used as input, such as particle hit momentum
$p_{x}$, show percent-level fluctuations and very good agreement.
Derived quantities, e.g.\ global coordinates, fluctuate to up to \qty{5}{\%}
but overall show good agreement, with deviations getting larger towards
the tails.

The track reconstruction efficiency is first compared between the original
and rounded \textsc{Geant4} samples. It drops for about \qty{2}{\%} for
the rounded case, indicating that the chosen quantisation procedure used
to tokenise hit data is not precise enough.

\begin{table}[t]
  \centering
  \caption{Track seeding and fitting efficiencies comparison between
    the \textsc{Geant4} simulation and simulation using neural networks
    for the models with full $\phi$ detector coverage}%
  \label{tab:efficiency}
  \begin{tabular*}{\columnwidth}{@{\extracolsep\fill}llcc}  
    \toprule
    Efficiency                  & Seeding        & Fitting        \\
    \midrule
    \textsc{Geant4} (reference) & \qty{99.9}{\%} & \qty{99.9}{\%} \\
    \textsc{Geant4} (rounded)   & \qty{99.9}{\%} & \qty{98.1}{\%} \\
    Transformer (\qty{11.2}{M}) & \qty{99.4}{\%} & \qty{94.9}{\%} \\
    Transformer (\qty{35.0}{M}) & \qty{99.7}{\%} & \qty{96.3}{\%} \\
    \bottomrule
  \end{tabular*}
\end{table}

The smallest benchmark model can reach comparable tracking performance to the
rounded \textsc{Geant4}.
The \qty{11.2}{M} parameter model can reach \qty{94.9}{\%} tracking efficiency
with only slight seeding efficiency drop. Increasing the layer dimensions by
a factor of two improves both seeding and tracking efficiencies, but they
are still lower than the reference sample.
The technical efficiencies are summarised in \cref{tab:efficiency}.

\begin{figure*}[htbp]
  \centering
  \includegraphics[width=0.9\columnwidth]{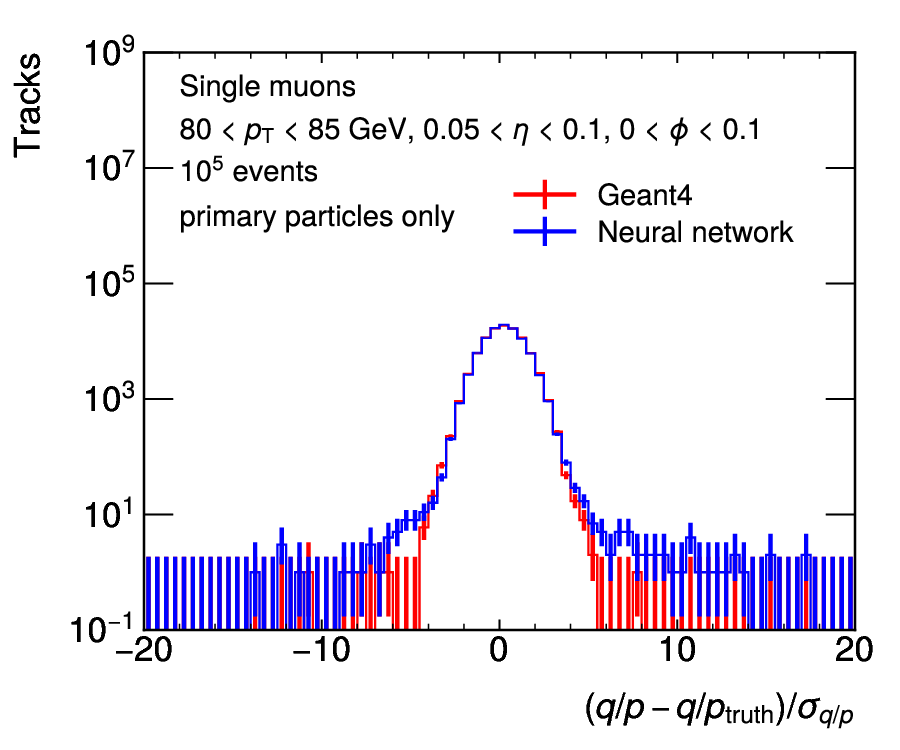}
  \quad
  \includegraphics[width=0.9\columnwidth]{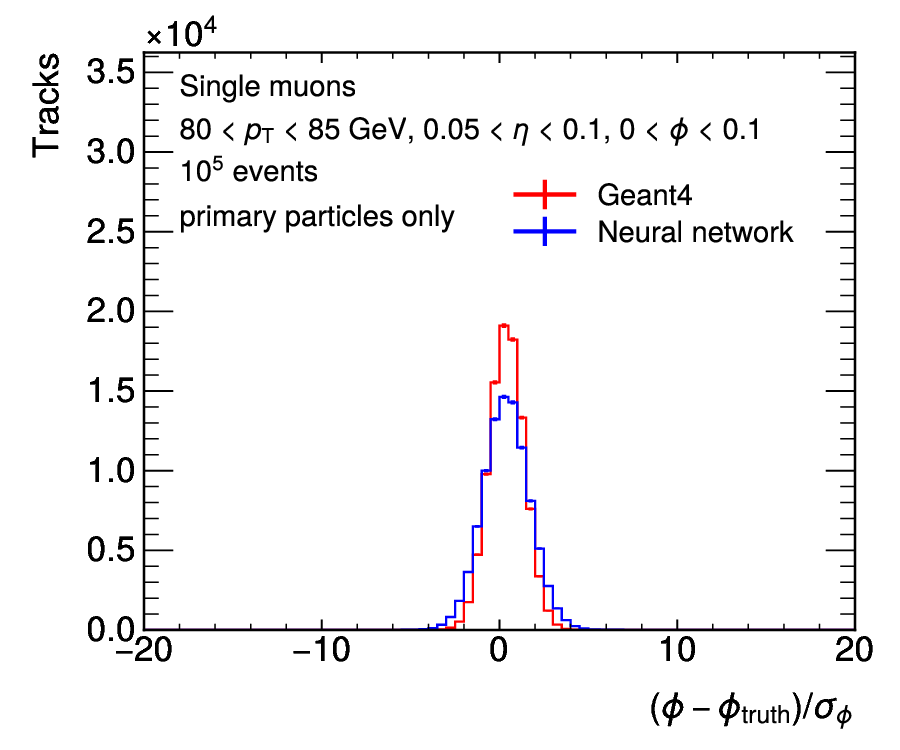}
  \caption{Comparisons of simulated track properties of single $\mu^{-}$
    particles reconstructed from rounded \textsc{Geant4} simulation output (red)
    and the neural network (blue).
    Track $q/p$ pull (left) and track $\phi$ pull (right) are shown}%
  \label{fig:results:tracks:mu}
\end{figure*}

\begin{figure*}[htbp]
  \centering
  \includegraphics[width=0.9\columnwidth]{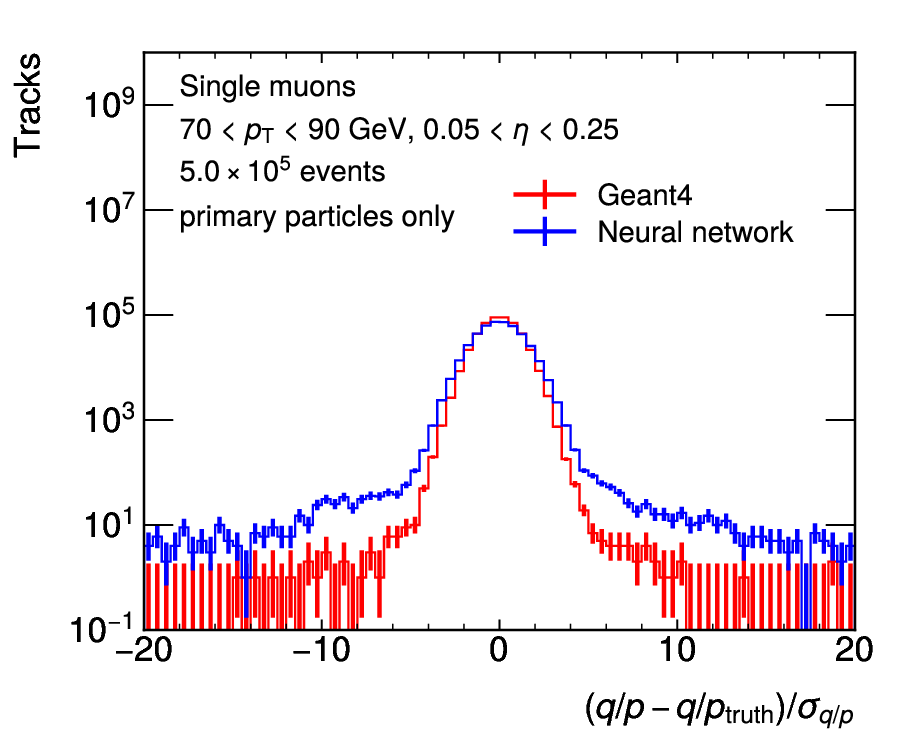}
  \quad
  \includegraphics[width=0.9\columnwidth]{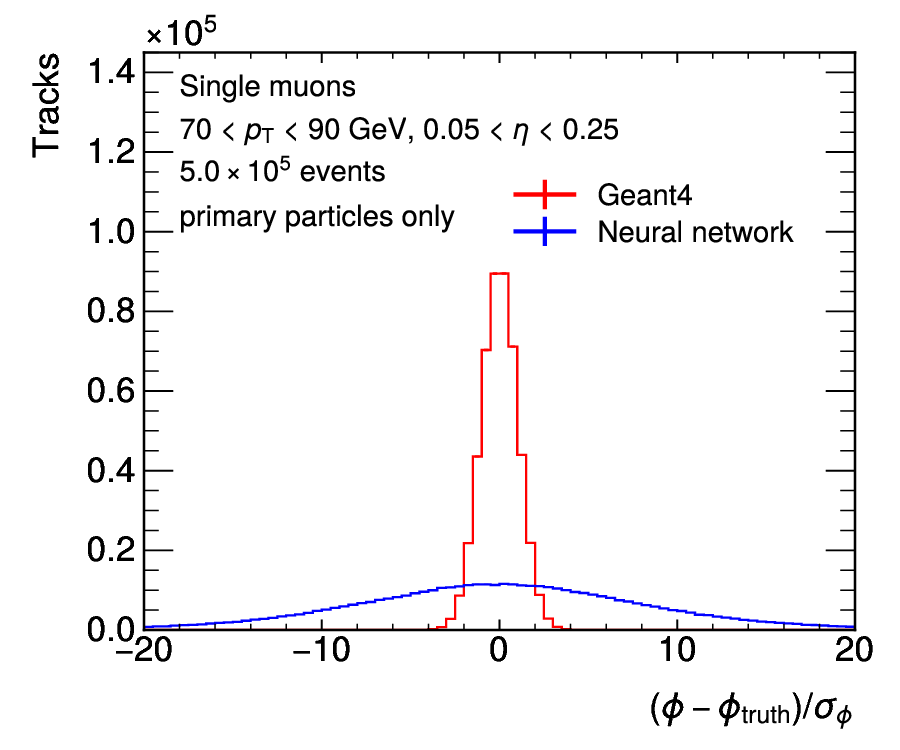}
  \caption{Comparisons of simulated track properties of single $\mu^{\pm}$
    particles reconstructed from rounded \textsc{Geant4} simulation output (red)
    and the neural network (blue).
    Track $q/p$ pull (left) and track $\phi$ pull (right) are shown}%
  \label{fig:results:tracks:mu_large}
\end{figure*}

\begin{figure*}[htbp]
  \centering
  \includegraphics[width=0.9\columnwidth]{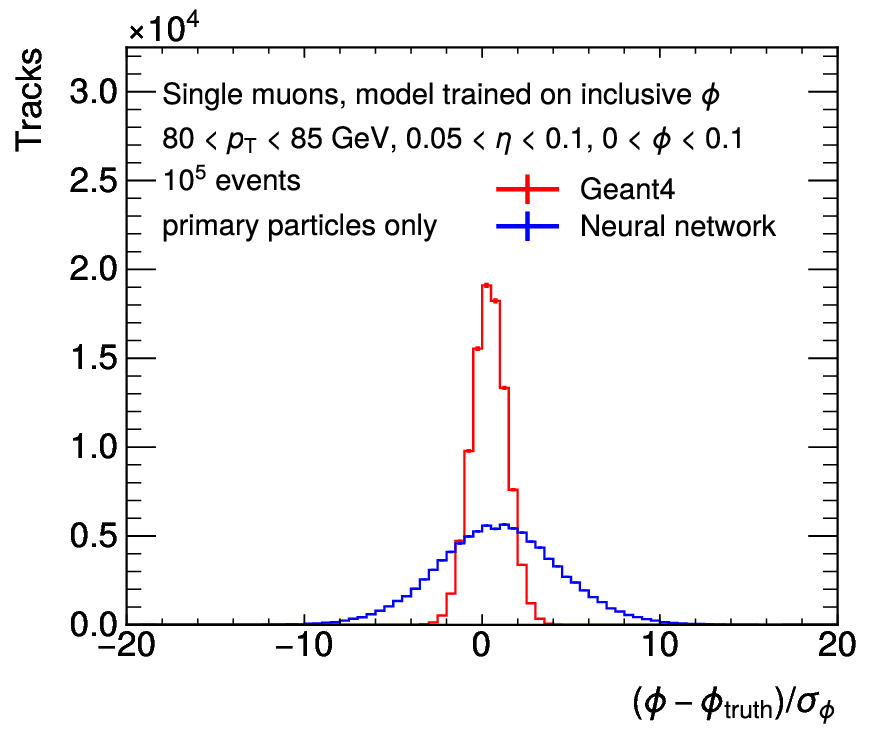}
  \quad
  \includegraphics[width=0.9\columnwidth]{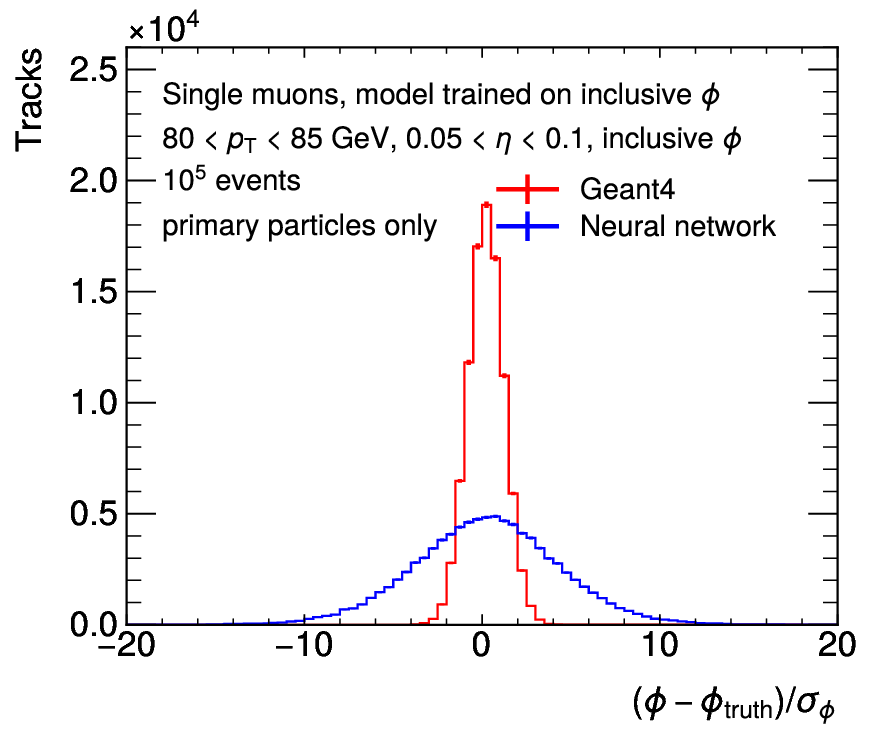}
  \caption{Comparisons of simulated track $\phi$ of single $\mu^{\pm}$
    particles reconstructed from rounded \textsc{Geant4} simulation output (red)
    and the neural network (blue), trained on the extended dataset without
    $\phi$ selection applied.
    Neural network is evaluated on the nominal (left) and extended (right) datasets}%
  \label{fig:results:tracks:phicut}
\end{figure*}

\Cref{fig:results:tracks:mu,fig:results:tracks:mu_large} show pulls of two
representative track properties, track $q/p$ and track $\phi$, for smaller and
larger muon datasets, respectively. The first is well modelled in both
cases with a slightly wider distribution and longer tails, especially for the
larger model. The $\phi$ coordinate does not learn well when the full $2\pi$
coverage of the detector is included. The large pull mainly comes from the
incorrectly reconstructed angle, the resolutions are comparable between
the \textsc{Geant4} and the transformer-generated inputs.
There is no significant improvement in the $\phi$ coordinate modelling when
increasing the number of parameters of the model.

As the angle $\phi$ is not modelled well, an additional dedicated test model was
trained on a benchmark dataset where the $\phi$ cut was removed.
Otherwise both the model configuration and sample statistics are the same
as with the nominal smaller muon dataset.
Performance was evaluated both on the nominal and the extended dataset.
\Cref{fig:results:tracks:phicut} shows pull of track $\phi$ using the model
trained on the extended dataset and evaluated on both of them, with additional
distributions in \cref{app:more-reco}.
While $\phi$ modelling is slightly better than the larger model
the pull distribution width is wider than the one from \textsc{Geant4}.
The tracking performance is very similar for both the nominal and the extended
dataset showing that the model is not sensitive to the specific set of events.
The conclusion is that with the increased phase space the model does not
learn the probabilities as accurately as with a more targeted selection.

\subsection{Single Electrons and Single Pions}

Electrons undergo much more significant bremsstrahlung and pair production
processes. While for this study secondary particles are not considered,
those additional processes can significantly alter the direction of the particle
or the particle momentum, compared to muons.

The same benchmark model is used for electrons as for muons.
The larger variation in the particle behaviour is confirmed by 1.4 times
larger token dictionary size. Implicitly this also makes the model about
\qty{10}{\%} larger.

\begin{figure}[b]
  \centering
  \includegraphics[width=0.9\columnwidth]{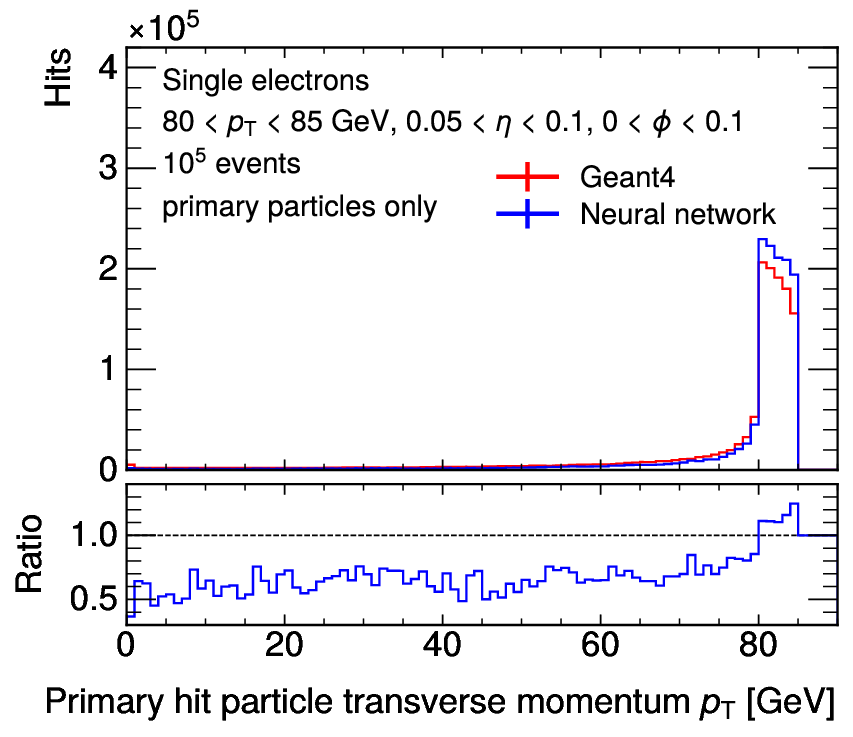}
  \caption{Comparison of simulated hit transverse momentum of single $e^{-}$
    particles simulated with \textsc{Geant4} (red) and the neural
    network (blue)}\label{fig:results:hits:el}
\end{figure}

\begin{figure}[htbp]
  \centering
  \includegraphics[width=0.9\columnwidth]{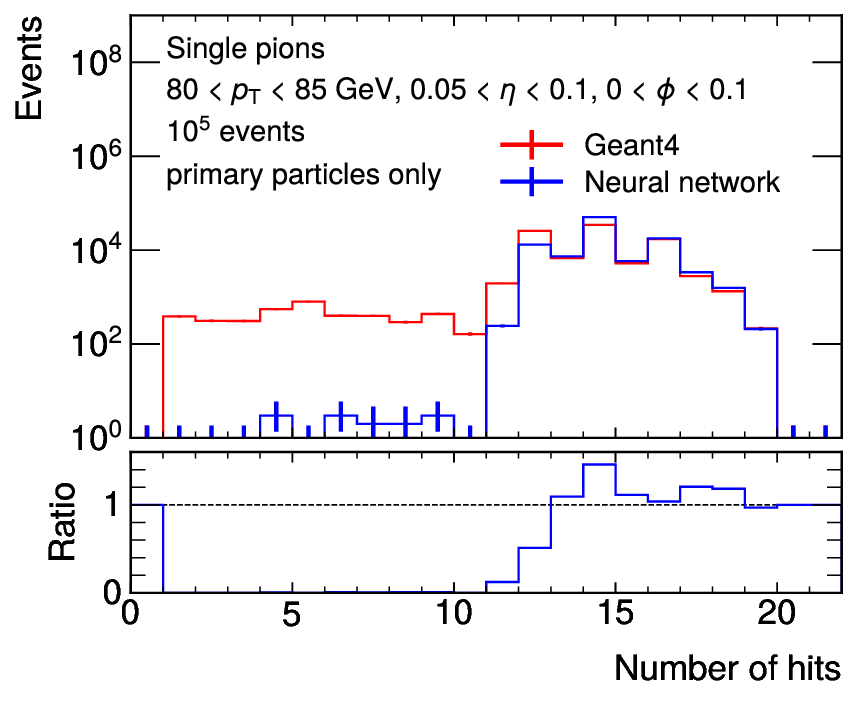}
  \caption{Comparison of number of simulated hits of single $\pi^{+}$
    particles simulated with \textsc{Geant4} (red) and the neural
    network (blue)}\label{fig:results:hits:pi}
\end{figure}

While the overall transformer model performance for electrons is similar to
the muon one, the momentum modelling is not sufficient. As seen in
\cref{fig:results:hits:el}, there is a large fraction of events that
have a too large transverse momentum. The model does learn the shape of the
momentum tail but is biased towards smaller changes.

This also translates to reconstructed tracks. Overall track quality is
still good and comparable with muons but the differences in particle momenta
at hit level also propagate to tracks.
Pion lifetime is about 100 times shorter than muons so they may decay already
inside the tracker system. This happens rarely, yielding a flat distribution
for lower number of hits, displayed in \cref{fig:results:hits:pi}.
Again this low-probability process is not properly learned and in the majority
of cases the ML-generated pions never decay in the detector.
This results in a too high tracking efficiency but tracks are in general
of good quality.
Representative electron and pion track properties distributions are available
in \cref{app:more-reco}.

\subsection{Computing Performance}

\begin{table}[b]
  \centering
  \caption{Comparison of inference speeds between CPUs and GPUs using standard
    floating-point precision for two model sizes per \num{10000} simulated tracks.
    \textsc{Geant4} simulation speed is provided as a reference.
    Total elapsed real time is measured}%
  \label{tab:speed:large}
  \begin{tabular*}{\columnwidth}{@{\extracolsep\fill}llcc}  
    \toprule
    \multicolumn{2}{@{\extracolsep\fill}l}{Compute type} & \qty{11.2}{M} & \qty{35.0}{M} \\
    \midrule
    CPU & 24 cores, AMD Zen 2    & \qty{35}{\minute}   & \qty{80}{\minute}   \\
    CPU & 24 cores, AMD Zen 4    & \qty{17}{\minute}   & \qty{37}{\minute}   \\
    \midrule
    GPU & Nvidia A100 \qty{40}{GB} PCIe & \qty{16.4}{\second} & \qty{36.0}{\second} \\
    GPU & Nvidia H100 \qty{80}{GB} PCIe & \qty{11.4}{\second} & \qty{21.4}{\second} \\
    GPU & Nvidia H100 \qty{80}{GB} SXM  & \qty{8.0}{\second}  & \qty{13.9}{\second} \\
    \midrule
    \textsc{Geant4} & 24 cores, AMD Zen 2 & \multicolumn{2}{c}{\qty{35}{\second}} \\
    \textsc{Geant4} & 24 cores, AMD Zen 4 & \multicolumn{2}{c}{\qty{17}{\second}} \\
    \bottomrule
  \end{tabular*}
\end{table}

The transformer-based simulation is trained on the \textsc{Geant4} simulated
hits. For production use it is only sensible if the inference is as fast
or faster than the simulation it bases on. \Cref{tab:speed:large} compares
the inference speeds for the larger, $\phi$-inclusive models.
They are compared between several GPUs used on high performance computers (HPCs)
and the standard CPUs used on same HPCs. \textsc{Geant4} simulation speeds
are shown for reference. GPU inference is comparable in speed to
the \textsc{Geant4} simulation running on same-generation CPUs.
It scales about linearly with model size meaning that larger models with better
tracking performance are also slower. CPU speeds are several orders of magnitude
slower and are not suitable as a replacement of the current simulation.
Due to the much higher costs of high-performance GPUs, the current performance
is not sufficient for general production use.

\begin{table}[t]
  \centering
  \caption{Training and inference speeds on various Nvidia graphic accelerators,
    ordered by age and memory,
    for two different computational precisions, \texttt{fp32} and \texttt{bf16}.
    Total elapsed real time is measured}%
  \label{tab:speed:small}
  \begin{tabular*}{\columnwidth}{@{\extracolsep\fill}lccc@{}c@{\extracolsep\fill}}  
    \toprule
    & \multicolumn{2}{c}{Training}    & \multicolumn{2}{c}{Inference}                                          \\
    & \multicolumn{2}{c}{[s / epoch]} & \multicolumn{2}{c}{[s / \num{10000} tracks]}                                 \\
    Precision      & \texttt{fp32}                   & \texttt{bf16}                          & \texttt{fp32} & \texttt{bf16} \\
    \midrule
    A100 \qty{40}{GB} SXM  & 75                              & 50                                     & 13.0          & 9.34          \\
    A100 \qty{80}{GB} PCIe & 77                              & 52                                     & 15.1          & 11.1          \\
    H100 \qty{80}{GB} SXM  & 38                              & 26                                     & 6.78          & 5.21          \\
    GH200 \qty{96}{GB}     & 34                              & 23                                     & 6.23          & 5.68          \\
    B200 \qty{180}{GB} SXM & 31                              & 21                                     & 4.54          & 4.28          \\
    \bottomrule
  \end{tabular*}
\end{table}

The smaller benchmark model trained on muons is also evaluated on various
Nvidia GPUs, shown in \cref{tab:speed:small}. Training and inference speeds
are measured for two different precisions (\texttt{fp32} and \texttt{bf16})
on GPUs with three different connections, the consumer PCIe, the higher
bandwidth SXM server connection, and the unified memory setup in the GH200 chip.
The jumps between generations are noticeable (A100 vs H100 and H100 vs B200),
especially for inference. The speed up is not that noticeable with the newest
B200 model as the benchmark dataset is not as large and with \qty{180}{GB}
of memory the framework overhead becomes noticeable.

Half-precision computations speed-up the training for about one third
without any physics performance change. Similar speed-up is observed
for inference with older GPUs, but it is less noticeable with the newer ones.
Newer hardware will allow training and usage of larger models
that are expected to also yield better physics performance.

\section{Conclusions}\label{sec:conclusions}

A novel method to simulate silicon tracking detectors using deep neural
networks, namely the transformer model, has been developed. Good hit-level
physics performance is achieved for muons, but worse for electrons and pions as
low-probability processes are not modelled well.
While small benchmark samples can achieve comparable tracking performance
as the standard \textsc{Geant4}-based simulation, moving to larger
production-like models has its limitations.

Transformer model's biggest drawback for physics simulation is its discrete
nature when used as a generative model.
Tokenisation needs to be done in a careful way to keep the token
space small but the simulation sufficiently accurate. Detector aware rounding
and tokenisation will have to be performed.

The second drawback is the iterative nature of the model. The array of hits
can not be generated at once or per layer as it can usually be done for
calorimeters, but each feature is generated in a separate step.
While this ensures full correlations and in principle high accuracy
it also significantly slows down the inference due to the slow attention
mechanism, which is especially noticeable on CPUs.
Inference speeds on GPUs need to become significantly faster to compensate
for much higher infrastructure costs of such computing resources.

The benchmark models can generate hits resulting in high-quality tracks.
Once the phase-space becomes larger the $\phi$ coordinate becomes
harder to simulate. Even in cases when low-probability processes occur
too infrequently the quality of the tracks is preserved.
Modelling of such processes could be improved with weights but that would
reduce the statistical power of the simulated samples and benefits of
faster simulation would be lost.

Future research can expand to secondary particles production,
that could be represented with additional elements of the sequence.
Transformers have proven to be successful also in modelling physics
processes, but are currently limited by their needed size and consequently
inference speed, so other sequence-based machine learning architectures
will have to be evaluated.
We believe the observed limitations can be overcome with more
sophisticated model implementations.

\backmatter

\bmhead{Acknowledgements}
This publication is co-funded by the European Union's Horizon
Europe research and innovation programme under the Marie Sk\l{}odowska-Curie
COFUND Postdoctoral Programme grant agreement
No. 101081355 --- SMASH and by the Republic of Slovenia and
the European Union from the European Regional Development Fund.
Views and opinions expressed are however those of the authors only and do not
necessarily reflect those of the European Union or European Research Executive
Agency. Neither the European Union nor the granting authority can be held
responsible for them.

The authors acknowledge the financial support from the Slovenian Research Agency
through the research programme P1-0135 and research project J1-60028.  

Machine learning models trained and evaluated on the Vega and Arnes HPCs,
parts of the Slovenian National Supercomputing Network,
and the FRIDA research computing cluster, hosted by University of Ljubljana,
Faculty of Computer and Information Science.

\bmhead{Data availability}
Smaller datasets generated and analysed as part of this study are uploaded
to Zenodo and available at
\href{https://doi.org/10.5281/zenodo.17774551}{10.5281/zenodo.17774551}.
Larger samples can be reproduced using the openly available software.

\bmhead{Code availability}
The machine learning code used for this study, \texttt{SiliconAI},
is preserved on Zenodo and available at
\href{https://doi.org/10.5281/zenodo.17568416}{10.5281/zenodo.17568416}.
Validation code \texttt{SiliconAI Validator} is also preserved on Zenodo,
available at
\href{https://doi.org/10.5281/zenodo.17567586}{10.5281/zenodo.17567586}.

\bmhead{Open access}
Reproduction of this article or parts of it is allowed as specified in
the CC-BY-4.0 license.

\begin{appendices}
\crefalias{section}{appendix}

\section{Additional Distributions}\label{app:more-reco}

\Cref{fig:results:tracks:phicut} shows track $q/p$ using the model
trained on the extended dataset single muon dataset, without any $\phi$ cut,
and evaluated on both the nominal and the extended dataset.

\begin{figure*}[htbp]
  \centering
  \includegraphics[width=0.9\columnwidth]{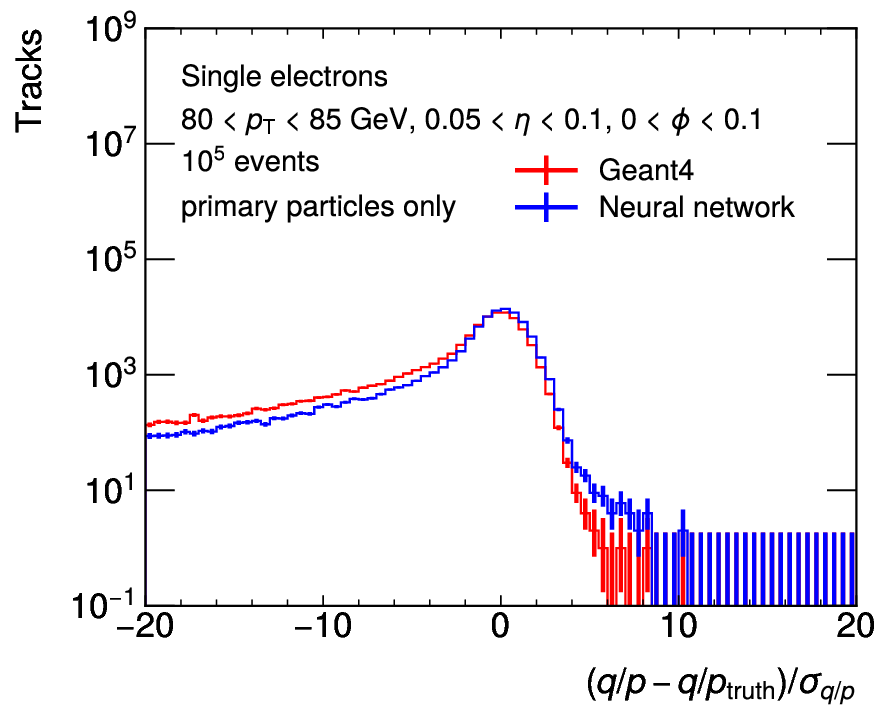}
  \quad
  \includegraphics[width=0.9\columnwidth]{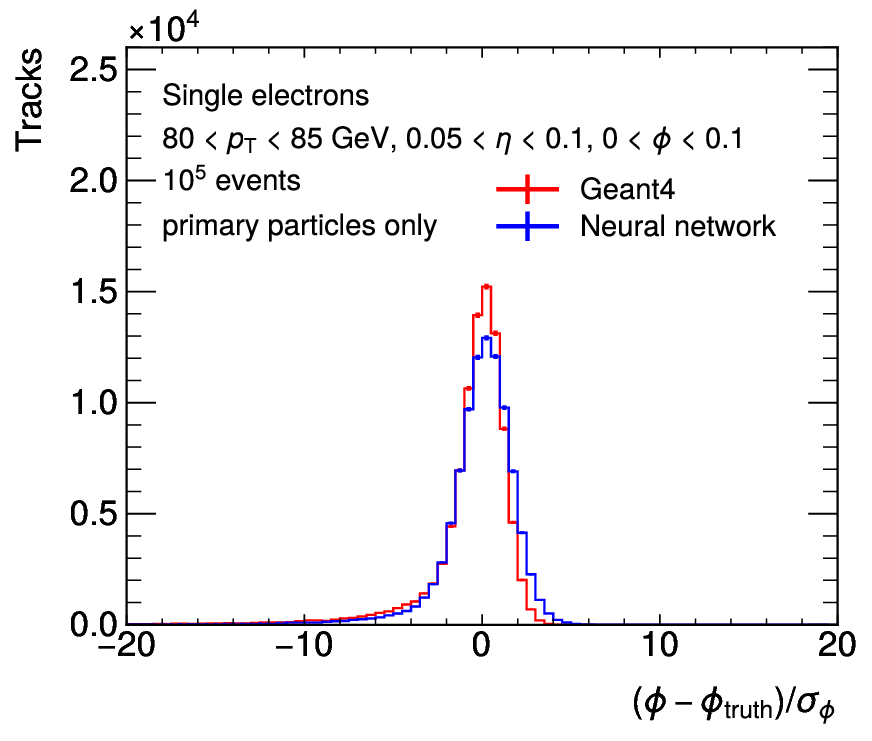}
  \caption{Comparisons of simulated track properties of single $e^{-}$
    particles reconstructed from rounded \textsc{Geant4} simulation output (red)
    and the neural network (blue).
    Track $q/p$ pull (left) and track $\phi$ pull (right) are shown}%
  \label{fig:results:tracks:el}
\end{figure*}

\begin{figure*}[htbp]
  \centering
  \includegraphics[width=0.9\columnwidth]{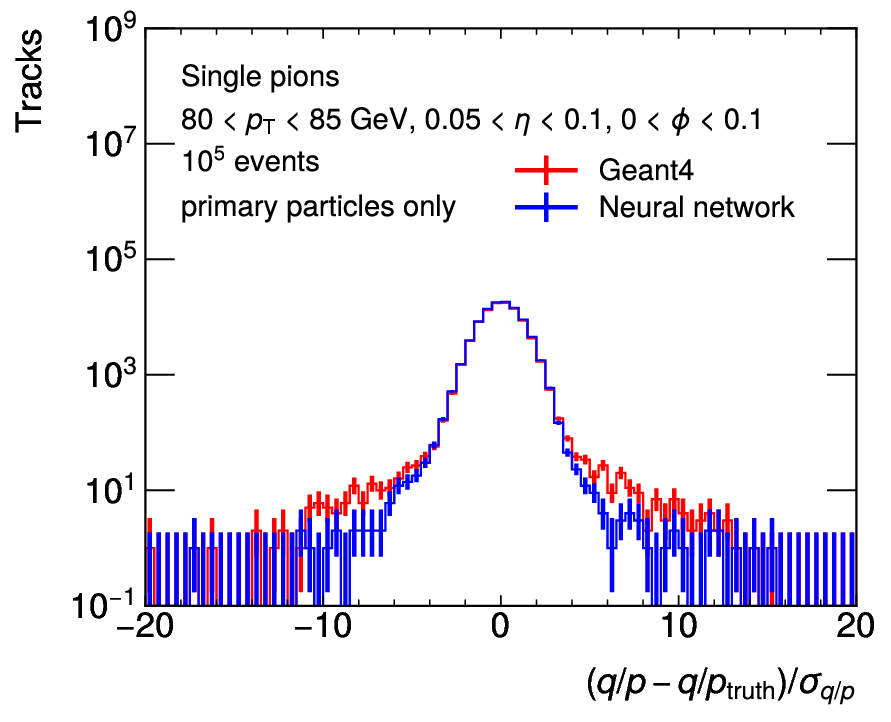}
  \quad
  \includegraphics[width=0.9\columnwidth]{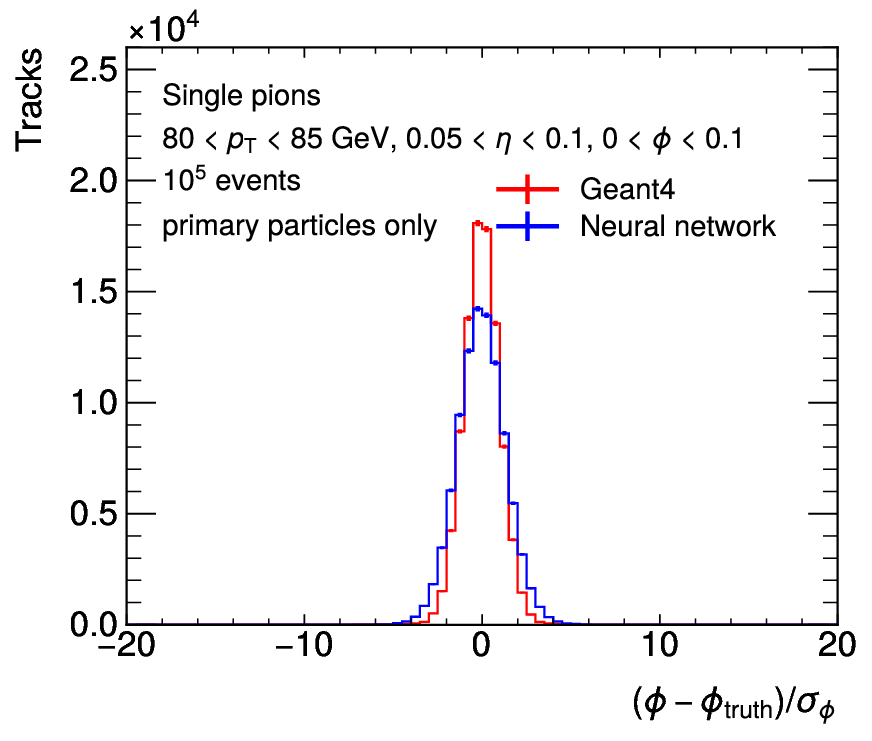}
  \caption{Comparisons of simulated track properties of single $\pi^{+}$
    particles reconstructed from rounded \textsc{Geant4} simulation output (red)
    and the neural network (blue).
    Track $q/p$ pull (left) and track $\phi$ pull (right) are shown}%
  \label{fig:results:tracks:pi}
\end{figure*}

\Cref{fig:results:tracks:el,fig:results:tracks:pi} show pulls of two
representative track properties, track $q/p$ and track $\phi$,
for electron and pion datasets, respectively.
They complement and are intended to be directly compared with
\cref{fig:results:tracks:mu}.

\begin{figure*}[htbp]
  \centering
  \includegraphics[width=0.9\columnwidth]{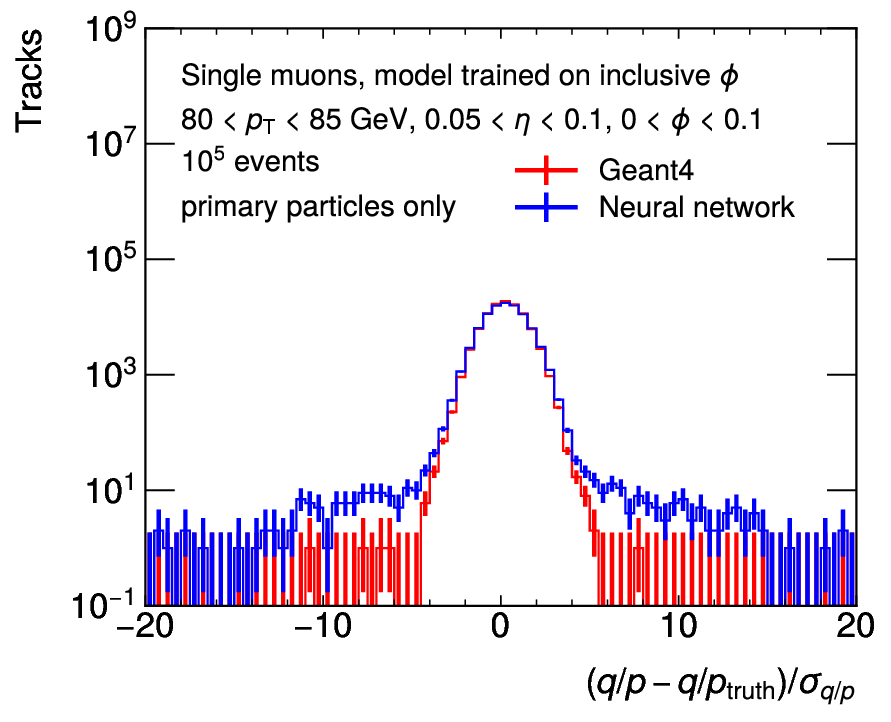}
  \quad
  \includegraphics[width=0.9\columnwidth]{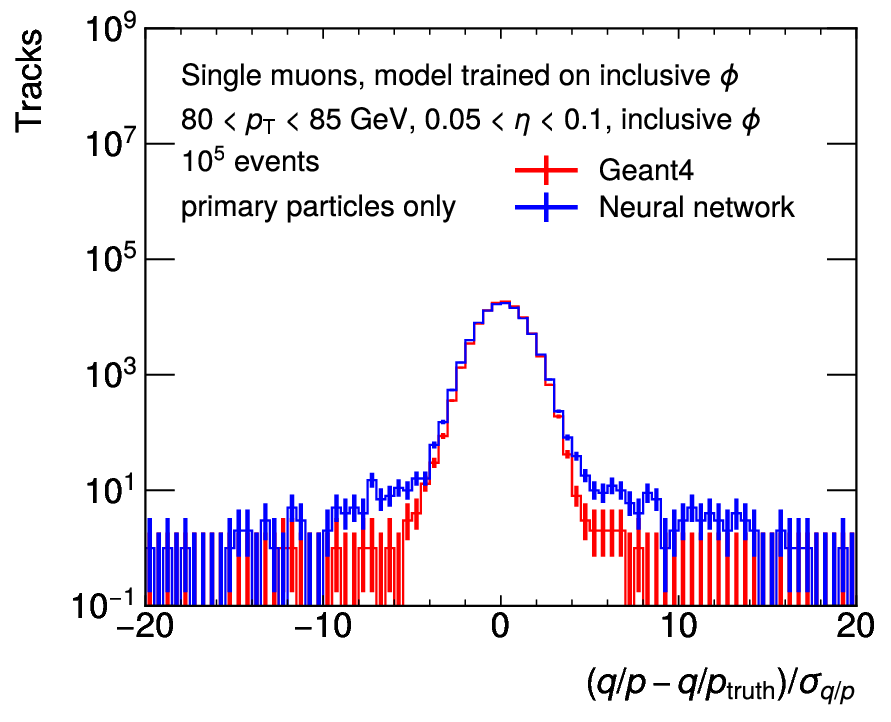}
  \caption{Comparisons of simulated track $q/p$ of single $\mu^{\pm}$
    particles reconstructed from rounded \textsc{Geant4} simulation output (red)
    and the neural network (blue), trained on the extended dataset without
    $\phi$ selection applied.
    Neural network is evaluated on the nominal (left) and extended (right) datasets}%
  \label{fig:results:tracks:phicut_qp}
\end{figure*}

\end{appendices}

\bibliography{sn-bibliography}

\end{document}